\documentclass[twocolumn,showpacs,preprintnumbers,amsmath,amssymb,prb]{revtex4-1}
\usepackage{graphicx}% Include figure files
\usepackage{dcolumn}% Align table columns on decimal point
\usepackage{bm}% bold math
\usepackage{color}
\usepackage{physics}
\usepackage{amsmath}
\usepackage[dvipsnames]{xcolor}

\newcommand{\ii}{{\rm i}}

\def\gsim{\lower.35em\hbox{$\stackrel{\textstyle>}{\textstyle\sim}$}}

\begin{document}

\title{Linear response of twisted bilayer graphene:  continuum vs. tight-binding models}

\author{T. Stauber$^{1}$, T. Low$^2$, and G. G\'omez-Santos$^3$, }

\affiliation{$^{1}$ Departamento de Teor\'{\i}a y Simulaci\'on de Materiales,
Instituto de Ciencia de Materiales de Madrid, CSIC, E-28049 Madrid, Spain\\
$^{2}$ Department of Electrical \& Computer Engineering, University of Minnesota, Minneapolis, Minnesota 55455, USA\\
$^3$Departamento de F\'{\i}sica de la Materia Condensada, Instituto Nicol\'as Cabrera and Condensed
Matter Physics Center (IFIMAC), Universidad Aut\'onoma de Madrid, E-28049 Madrid, Spain}
\date{\today}

\begin{abstract} 
We present  a linear response calculation for twisted bilayer graphene. The calculation is performed for both the continuum and tight-binding models, with the aim of assessing the validity of the former. All qualitatively important features previously reported by us [T. Stauber et al. Phys. Rev. Lett. {\bf 120}, 046801 (2018)] for the Drude matrix in the continuum model are also present in the tight-binding calculation, with increasing quantitative agreement  for decreasing twist angle. These features include the  chiral longitudinal magnetic moment associated with  plasmonic modes, and the anomalous counterflow around the neutrality point, better interpreted as a paramagnetic response.  We have addressed the differences between Drude and equilibrium response, and shown that orbital paramagnetism is the equilibrium response to a parallel magnetic field over a substantial doping region around the neutrality point. Chirality also makes the equilibrium response to exhibit a non trivial current structure associated with the non-vertical character of interlayer bonds in the tight-binding calculation. 
\end{abstract}

\maketitle

%%%%%%%%%%%%%%%%%%%%%%%%%%%%%%%%%%%%%%%%%%%%%%%%%%%%%%%%%%%%%
%  SECTION INTRODUCTION
%%%%%%%%%%%%%%%%%%%%%%%%%%%%%%%%%%%%%%%%%%%%%%%%%%%%%%%%%%%%%
\section{INTRODUCTION}

Chiral molecules, ubiquitous in natural and synthetic organic chemistry,  have long been the subject of much  attention and  used in many applications.\cite{Barron04} More recently,  plasmonic metamaterials and other artificial nanostructures  with chiral  capabilities have also been implemented.\cite{Tang11,Zhao17,Guerrero11,Shen13,Hentschel17} The design of atomically thin two dimensional van der Waals materials\cite{Geim13} has enlarged the list of artificial optically active material significantly, i.e., any combination of misaligned van der Waals materials should lead to circular dichroism which can further be enhanced by increasing the number of twisted layers.\cite{Kim16} 

Twisted bilayer graphene is the most widely studied system among misaligned van der Waals structures. It is made of two graphene layers rotated by an arbitrary angle with respect to each other.\cite{Li10,Schmidt10,Brihuega12,Dean13,Havener14,Schmidt14,Patel15} Its non-interacting electronic structure mimics its geometry, with  two Dirac cones  displaced in the Brillouin zone by the twist angle.\cite{Lopes07,Bistritzer11} But correlation effects become important for filling factors close to the neutrality point,\cite{Kim17} leading to the opening of a Mott gap\cite{Cao18a} and to a superconducting phase\cite{Cao18b} that turns out to be tuneable.\cite{Yankowitz18} Also twisted structures consisting of other van der Waals materials such as MoS$_2$ have been investigated showing a modulated red shift of the excitonic gap.\cite{Liu14} Also in hetero-bilayers, interlayer excitons are long-lived\cite{Rivera15,Kunstmann18} and can be confined by the moir\'e lattice, potentially leading to quantum information applications.\cite{Tran18}

Twisted bilayer graphene (TBG) is a chiral material because its geometry is not parity invariant, with left- and right-handed copies corresponding to opposite twist angles. Indeed, TBG experimentally exhibits significant optical activity at finite frequencies corresponding to transitions  with strong interlayer hybridization around the $K$ and the $M$-point,\cite{Kim16} without the need of a magnetic field.\cite{Poumirol17}

The theoretical explanation of TBG optical activity has been considered in Refs. \cite{Kim16,Suarez17}. Motivated by the ever increasing sophistication of experimental transport results,  we have recently extended the calculation of TBG response  to zero frequencies\cite{Stauber18}, obtaining the Drude matrix where the excitation and response of each layer can be discriminated. Such a calculation, performed within the framework of the continuum model,  has unveiled potentially relevant results. These include, for instance,   the emergence of a longitudinal magnetic moment accompanying currents, such as those of intrinsic plasmons, endowing them with a chiral character. Also, we obtained counter intuitive behavior in a counterflow configuration, where opposing currents in each layer seem to flow opposite to their respective  electric field even at zero doping. All this might be interesting in view of manipulating the electronic properties of two-dimensional layered structures through their twist angle - so-called "twisttronics".\cite{Carr17}

This work is largely devoted to an assessment of the linear response validity of the continuum model of TBG. For this, the Drude weight, which is the key quantity in the dynamics of plasmons\cite{Chen12,Fei12} and which can also be obtained from transport meaurements,\cite{Yoon14} is calculated and shown that it needs to be extended to a Drude matrix. We then compare the predictions of a tight-binding model with those of its  continuum counterpart. This analysis is important because the continuum model or some variant of it will be needed if we ever want to  address the smallest angles within Bloch theory. For non-commensurate structures, novel techniques are needed.\cite{Massatt17,Cances17}

A further motivation for this study comes from the observation made in Ref. \cite{Suarez17} that, in explaining the experimentally observed circular dichroism, the continuum model is vulnerable to otherwise accepted approximations. 
The peculiar effects  obtained by us in the continuum model and, particularly those associated with chirality, are typically small. Given the possibility, however remote, that such behavior could be an artifact of the continuum model, we consider  its assessment against a tight-binding calculation as imperative. 

Although the numerical effort limits the tight-binding calculation to rather large angles,  as argued in ref. \cite{Lopes12}, the continuum model  by its very construction  should become a  better description of TBG for decreasing angles. Therefore, agreement in the nominally worst case of large angles becomes more relevant. The results to be presented later confirm that all qualitative features of the continuum calculation are indeed present in the tight-binding results, with quantitative agreement increasing with decreasing twist angle, as expected. The comparison will not be limited to the Drude matrix, but also the equilibrium response in the presence of a parallel magnetic field will be presented, where similar degree of agreement is found.

The paper is organized as follows. Section \ref{TB} present the tight-binding model, its linear response formalism, and a physical discussion of the response terms, largely valid also for the continuum case. Section \ref{CM} contains a brief account of the continuum model and its response, already presented in Ref.\cite{Stauber18}, to make the work self-contained. Section \ref{RS} presents the main results of this work together with their physical discussion,  both for the Drude and equilibrium cases.  Section \ref{SM} summarizes the main findings. Three appendices are included with details of the tight-binding Hamiltonian and the linear response calculation.

%{\it Hamiltonian.}
\section{Tight-binding Model}\label{TB}
\subsection{Geometry and Hamiltonian}
We consider two parallel graphene layers with lattice constant $a_g=2.46 \, \mathring{\text{A}} $, separated along the $z$ axis by a distance $a=3.5 \, \mathring{\text{A}}$,  with the second layer rotated with respect to  a 
$A_1B_2$ stacking point by an angle $\theta_i $, with $\cos(\theta_i)=1-\tfrac{1}{2(3i^2 + 3i + 1)}$ for integer $i$, so that a commensurate superstructure results. The Hamiltonian can be written as 
\begin{equation}\label{Hamiltonian}
\mathcal{H}_0 =\mathcal{H}_{1} + \mathcal{H}_{2}+ \mathcal{H}_{inter}
,\end{equation}
where $\mathcal{H}_{1(2)} $ corresponds to the intralayer Hamiltonian, described by a single nearest-neighbor tight-binding hopping integral $t$, with $t=3 \,\text{eV}$.  $\mathcal{H}_{inter}$ describes the interlayer hopping, and it is given by 
\begin{equation}\label{Hinter}
\mathcal{H}_{inter} =\sum_{i\in1,j\in2} V(d_{ij}) \; c_{i}^{\dagger} \, c_{j} + H.\; c.
,\end{equation}
where $V(d_{ij}) $ only depends  on the distance between orbitals, 
% $ d_{ij}=\sqrt{r_{ij}^2+a^2}$, where $r_{ij} $ is the in-plane separation,  
 so that the analysis of ref.\cite{Bistritzer11} applies. The details of $V(d_{ij}) $ are provided in the appendix \ref{inter}, suffice it to say here that the largest interlayer hopping integral is taken to be around 
16 percent\cite{Li10,Moon13} of the intralayer $t$.

\subsection{Linear Response}
We will only consider fields and currents parallel to the planes. Furthermore, we will temporarily restrict our attention to horizontally homogeneous fields while allowing spatial variation along the stacking direction, so that only the $\bm q=0 $ Fourier component survives.  Under these conditions, the linearly perturbed Hamiltonian is
\begin{equation}\label{Hperturb}
\mathcal{H} =\mathcal{H}_0 + \mathcal{V}
,\end{equation}
with 
\begin{equation}\label{V}
\mathcal{V} = -S \; [ \bm j_p^{(1)} \cdot \bm A^{(1)} +\bm j_p^{(2)} \cdot \bm A^{(2)} +\bm j_p^{(inter)}  \cdot \bm A^{(inter)} ] 
,\end{equation}
with layer surface $S$. $ \bm A^{(1,2)} $  are the fields at the graphene layers $1$ and $2$ , and $ \bm A^{inter} $  is that at the mid-point  between  graphene layers.     $\bm j_p^{(1,2,inter)} $ are the corresponding paramagnetic current operators, given explicitly 
in the appendix \ref{Lresponse}. Notice that $\bm j^{(inter)}$ accounts for the fact that, in the tight-binding model, a non-vertical interlayer bond can carry a parallel current. We will use the ordering
\begin{equation}\label{A}
\bm A =
\left[
\begin{array}{l}
\bm A^{(1)} \\  \bm A^{(2)} \\ \bm A^{(inter)}
\end{array} 
\right]
,\end{equation}
and
\begin{equation}\label{j}
\bm j=
\left[
\begin{array}{l}  \bm j^{(1)} \\  \bm j^{(2)} \\  \bm j^{(inter)} 
\end{array} 
\right] 
,\end{equation}
where $\bm j $ stands for the physical current $\bm j $, which includes a diamagnetic contributions, $\bm j_d$, so that 
\begin{equation}
\bm j = \bm j_p + \bm j_d
.\end{equation}

 The induced paramagnetic currents  can then be written as
\begin{equation}\label{chip}
\bm j_p = -\bm \chi_p \bm A
,\end{equation}
where a ground state average is implicit for the left-hand-side of Eq. \ref{chip}. The $6\times 6$   ($3$ currents $ \times$  $2$ components) tensor $ \bm \chi_p$ is forced by the symmetries of the problem to have the form:
\begin{equation}\label{chipbis}
\bm \chi_p = 
	\left[
	\begin{array}{cc|cc|cc}
	 \chi_0 & 0      &  \chi_1     & \chi_{xy} & \chi_{2}   & \chi'_{xy} \\
	  0     & \chi_0 & -\chi_{xy}  & \chi_{1}  & -\chi'_{xy} & \chi_2      \\
	\hline
	  \chi_1  & -\chi_{xy}   &  \chi_0 & 0      &  \chi_{2}   & -\chi'_{xy} \\
	 \chi_{xy}  & \chi_{1} &   0     & \chi_0 &    \chi'_{xy} & \chi_2    \\
       \hline
         \chi_{2}   & -\chi'_{xy} & \chi_{2}   & \chi'_{xy} & \chi_i & 0 \\  
         \chi'_{xy} & \chi_2      & -\chi'_{xy} & \chi_2    &  0     & \chi_i
	\end{array}
	\right]
,\end{equation}
where the linear response  calculation of non-zero entries is described in the appendix \ref{Lresponse}.
  
Likewise, the diamagnetic contribution can be written as
\begin{equation}\label{chid}
\bm j_d = -\bm \chi_d \bm A
,\end{equation}
where, again, symmetries reduce  the tensor $\bm \chi_d  $ to the diagonal form
\begin{equation}\label{chidbis}
\bm \chi_d = 
	\left[
	\begin{array}{cc|cc|cc}
	 \chi_{d0} & 0      &  0     & 0 & 0   & 0 \\
	  0     & \chi_{d0} & 0  & 0  & 0 & 0     \\
	\hline
	  0  & 0   &  \chi_{d0} & 0      &  0  & 0 \\
	 0  & 0 &   0     & \chi_{d0} &   0 & 0    \\
       \hline
         0  & 0 & 0   & 0 & \chi_{di} & 0 \\  
         0& 0     & 0 & 0   &  0     & \chi_{di}
	\end{array}
	\right]
,\end{equation}
with the calculation of non-zero entries explicited in the appendix \ref{Lresponse}.

\subsection{Drude matrix}

The physical current can be written as 
\begin{equation}
\bm j = - \bm \chi \bm A
,\end{equation}
where $ \bm \chi = \bm \chi_p + \bm \chi_d $. The expressions given in  Eqs. \ref{chipbis} and \ref{chidbis} correspond to the $\bm q=0$ but arbitrary frequency, so that all entries are frequency functions. Indeed, the chiral entries $\chi_{xy}(\omega)$ and $\chi'_{xy}(\omega) $ are responsible for the experimentally observed circular dichroism at optical frequencies. As in ref.\cite{Stauber18}, we will be concerned with the $  \omega \to 0$ limit, which physically corresponds to the Drude weight, here promoted to a Drude matrix. Therefore, we define the Drude matrix as 
\begin{equation}\label{Drude}
\bm D = \lim_{\omega \to 0} \bm \chi 
,\end{equation}
given explicitly by
\begin{equation}\label{Drudebis}
\bm D = 
	\left[
	\begin{array}{cc|cc|cc}
	 D_0 & 0      &  D_1     & D_{xy} & D_{2}   & D'_{xy} \\
	  0     & D_0 & -D_{xy}  & D_{1}  & -D'_{xy} & D_2      \\
	\hline
	  D_1  & -D_{xy}   &  D_0 & 0      &  D_{2}   & -D'_{xy} \\
	 D_{xy}  & D_{1} &   0     & D_0 &    D'_{xy} & D_2    \\
       \hline
         D_{2}   & -D'_{xy} & D_{2}   & D'_{xy} & D_i & 0 \\  
         D'_{xy} & D_2      & -D'_{xy} & D_2    &  0     & D_i
	\end{array}
	\right]
,\end{equation}
where, for instance,  $ D_0 = \lim_{\omega \to 0} [\chi_0(\omega)+\chi_{d0}] $   and similarly the remaining entries.

The Drude matrix is essentially a dynamical concept: it measures the system density of inertia (inverse mass) resisting the (slow) acceleration of a currents by electric fields.   This is best seen by writing the electric field as $ \bm E = \ii \omega \bm A $ and rewriting the response as 
\begin{equation}\label{j1}
- \ii \omega \, \bm j  =  \bm \chi \, \bm E
,\end{equation}
which, upon restoring the time, is equivalent to 
\begin{equation}\label{j2}
 \partial_t \bm j  =  \bm D \, \bm E
,\end{equation}
for slow variations. Introducing a phenomenological scalar dissipation $\tau$, Eqs. \ref{j1} and \ref{j2} are equivalent to a matrix generalization of the more familiar expression for the conductivity, $\bm \sigma = \tfrac{1}{-\ii \omega + \tau^{-1}} \bm D $.

\subsection{Physical interpretation}\label{physical}

  The Drude matrix  of Eq. \ref{Drudebis} provides the most complete information of the response for $\bm q=0 $ and $\omega \to 0 $, and we will present results for all entries later. But prior to that, it is convenient to adopt a slightly different view   in order to gain more physical insight. What follows is a generalization of out treatment of ref.\cite{Stauber18} to the full tight-binding case.  

Firstly, we can assume that  the field changes linearly between layers, correct to lowest order. Then the three perturbing fields can be written as 
\begin{equation}
\begin{split}
 & \bm  E^{(inter)} = \bm E_{\parallel} \\
 & \bm E^{(1)} = \bm E_{\parallel} +  (\bm E^{(1)} - \bm E^{(2)} ) / 2 \\
 & \bm E^{(2)} = \bm E_{\parallel} -  (\bm E^{(1)} - \bm E^{(2)} ) / 2 
,\end{split}
\end{equation}
so that the perturbation can be spelled in terms of the average parallel field, $\bm E_{\parallel} $, and its change across the bilayer,  
 $(\bm E^{(1)} - \bm E^{(2)} )$ , later related to the magnetic field.

Correspondingly, we will focus on the total current response, $ \bm j_T$, and its variation, $\bm j_m $,  
\begin{equation}
\begin{split}
 \bm j_T &= \bm j^{(1)} + \bm j^{(2)} + \bm j^{(inter)} \\
 \bm j_m  &= (\bm j^{(1)}-\bm j^{(2)})/2
.\end{split}
\end{equation}
Note that $\bm j_m $ will be non-zero if the layers are driven in opposite direction, the couterflow configuration considered in Ref.\cite{Bistritzer11} We will later  relate it to the magnetic moment, whereof the notation.

Using the Drude matrix in Eq. \ref{Drudebis} , one can show that the physical response can be cast in the form  of the following constitutive relations 
\begin{equation}
\begin{split}
 \partial_t \bm j_T  &=   D_T \, \bm E_{\parallel} + D_{chir} \, \bm{\hat{z}} \times (\bm E^{(2)} - \bm E^{(1)}) \\
 \partial_t \bm j_m  &= D_{chir} \, \bm{\hat{z}} \times \bm E_{\parallel} - \frac{D_{mag}}{2} (\bm E^{(1)} - \bm E^{(2)}) \label{constit0}
,\end{split}
\end{equation}
where we have introduced the {\em total} $(D_T)$, {\em chiral} $(D_{chir})$, and  {\em counterflow} or {\em magnetic}  $(D_{mag})$, Drude parameters, given by
\begin{align}
 & D_T  = 2 (D_0+D_1) + 4 D_2 + D_i \label{DT} \\
 & D_{chir}  = D_{xy} + D'_{xy} \label{Dchir} \\
 & D_{mag}  = D_1 - D_0 \label{Dmag}
.\end{align}

The magnetic language is introduced using  Maxwell equations  to   write
\begin{equation}
\bm{\hat{z}} \times (\bm E^{(2)} - \bm E^{(1)}) = - a \, \partial_t \,  \bm B_{\parallel}  
,\end{equation}
where $\bm B_{\parallel} $ is the parallel magnetic field.
Therefore, we can rewrite the constitutive relations as
\begin{equation}\label{constit}
\begin{split}
 \partial_t \bm j_T  &=   D_T \, \bm E_{\parallel} - a D_{chir}  \; \partial_t \bm B_{\parallel}   \\
 \partial_t \bm m_{\parallel}  &= a D_{chir}  \; \bm E_{\parallel} + \frac{a^2}{2} D_{mag} \; \partial_t \bm B_{\parallel}  
,\end{split}
\end{equation}
where the parallel magnetic moment density, $\bm m_{\parallel}   = a \, \bm j_m \times \bm{\hat{z}}$, has been introduced. 

Notice that, if only a magnetic field is present, one can drop the time derivatives,  leading to
\begin{align}%\label{adiabatic}
%\begin{split}
  \bm j_T  &=    - a D_{chir}  \;  \bm B_{\parallel}  \label{adiabaticj} \\
  \bm m_{\parallel}  &=   \frac{a^2}{2} D_{mag} \;  \bm B_{\parallel}  \label{adiabaticm}
%,\end{split}
.\end{align}
%
%
%valid for $\bm E_{\parallel}=0 $.
It  is important not to forget the dynamical meaning of the previous expression.  It is the adiabatic application of a magnetic field what results in a total parallel current and, perhaps less surprisingly, a magnetic moment. The associated currents  are produced by the transient electric fields, and the ideal dissipationless nature of the calculation makes those currents permanent. This has two consequences. Firstly,  the practical observation would require a dynamical measurement with  $ \omega \tau >> 1 $ , as stressed in our previous work\cite{Stauber18}. Secondly, even in the ideal dissipationless case, the current and magnetic moment of Eqs. \ref{adiabaticj} and \ref{adiabaticm} need not coincide with the equilibrium response in the presence of a magnetic field. This issue is treated in detail in section \ref{equilresp}. Let us mention that dissipationless counterflow at the neutrality point was also seen in the context of superfluid exciton flow, but only in the quantum Hall regime under the influence of a strong magnetic field in perpendicular sheet-direction.\cite{Liu17}

On symmetry grounds, Eq. \ref{adiabaticj} is allowed as both current and field have the same signature upon time reversal. On the other hand, current and field have opposite signature under parity reversal, and Eq. \ref{adiabaticj} would be forbidding for a parity invariant system. Of course,  lack of parity invariance is  precisely what chirality means and, therefore, Eq. \ref{adiabaticj} is allowed.

Finally, we consider the effect of the chiral terms on plasmons. Doped TBG, as graphene\cite{Chen12,Fei12,Yan13} or any 2d metal,  exhibits self-sustained charge oscillations\cite{Stauber13,Stauber16}. These can be obtained from the constitutive equations as shown in Ref.\cite{Stauber18}. Adapting that treatment to the present case, the plasmon  dispersion is given by $\omega_p(q) = \sqrt{\tfrac{D_T}{2 \epsilon_0} q }$, where the chiral terms do appear. 
Nevertheless, the chiral contributions add a transverse component to the plasmon current, given  by the following relation between electric and magnetic dipole oscillations:
\begin{equation}\label{constraint}
 {\bf \hat q}\cdot \bm m = a \frac{D_{chir}}{D_T}\; {\bf \hat q}\cdot \bm j_T
,\end{equation}
as is easily shown from the constitutive relations ignoring magnetic self-fields (instantaneous approximation).
 Therefore, the  plasmon carries total charge $\bm q \cdot \bm j_T \neq 0 $ and, by the constraint of Eq. \ref{constraint}, also carries a longitudinal magnetic moment, the hallmark of chiral excitations\cite{Rosenfeld1926,Barron04}. Thus, the finite value of the chiral Drude terms, $D_{xy} $ and $D'_{xy} $,  bestows plasmons with chiral character.

\subsection{Equilibrium response}\label{equilresp}

The Drude response, in spite of the limit $\omega \to 0$, is 
a dynamical magnitude, as already explained. Here we consider the true equilibrium response.
At the formal level,  equilibrium, $\bm \chi_{eq}$, and Drude responses to a vector potential  only differ in the order of limits,
\begin{align}
\bm \chi_{eq} &= \lim_{\bm q \to 0} \lim_{\omega \to 0} \bm \chi(\bm q,\omega)  \\
\bm D &= \lim_{\omega \to 0} \lim_{\bm q \to 0} \bm \chi(\bm q,\omega)  
,\end{align}
and writing the equilibrium response in the tight-binding case as
\begin{equation}\label{chieq}
\bm \chi_{eq} = 
	\left[
	\begin{array}{cc|cc|cc}
	 \tilde \chi_0 & 0      &  \tilde \chi_1     & \tilde \chi_{xy} & \tilde \chi_{2}   & \tilde \chi'_{xy} \\
	  0     & \tilde \chi_0 & -\tilde \chi_{xy}  & \tilde \chi_{1}  & -\tilde \chi'_{xy} & \tilde \chi_2      \\
	\hline
	  \tilde \chi_1  & -\tilde \chi_{xy}   &  \tilde \chi_0 & 0      &  \tilde \chi_{2}   & -\tilde \chi'_{xy} \\
	 \tilde \chi_{xy}  & \tilde \chi_{1} &   0     & \tilde \chi_0 &    \tilde \chi'_{xy} & \tilde \chi_2    \\
       \hline
         \tilde \chi_{2}   & -\tilde \chi'_{xy} & \tilde \chi_{2}   & \tilde \chi'_{xy} & \tilde \chi_i & 0 \\  
         \tilde \chi'_{xy} & \tilde \chi_2      & -\tilde \chi'_{xy} & \tilde \chi_2    &  0     & \tilde \chi_i
	\end{array}
	\right]
,\end{equation}
it is shown in the appendix \ref{appequilibrium} that each equilibrium entry only differs from the corresponding Drude one in a Fermi surface term whose calculation is there detailed.

In addition to the symmetries already considered in writing Eq. \ref{chieq}, {\em gauge invariance} imposes further constraints. The fact that a globally uniform vector potential, $\bm A^{(1)}=\bm A^{(2)}=\bm A^{(inter)} $, should have no physical consequences (currents), enforces the following relations among the equilibrium matrix entries:
\begin{align}
\tilde \chi_0 + \tilde \chi_1 + \tilde \chi_2 & = 0 \\
\tilde \chi_i + 2 \tilde \chi_2 & = 0 \\
\tilde \chi_{xy} + \chi'_{xy}   & = 0  \label{gaugetight}
.\end{align}
These consistency requirements have been verified in our calculation to numerical accuracy.

\section{Continuum  Model} \label{CM}
Here we just outline the basic points of the continuum description, referring the reader to references \cite{Lopes07,Bistritzer11,Lopes12} for details. 
The Hamiltonian is written as
\begin{align}%\label{Hamiltonian}
\mathcal{H } =\hbar v_F\sum_{\bm k,\alpha,\beta} [&c_{1,\bm k,\alpha}^{\dagger} \;\bm \tau_{\alpha\beta}^{-\theta/2} \cdot (\bm k + \frac{\Delta\bm K}{2}) \; c_{1,\bm k,\beta} \notag\\ 
  + \, &c_{2,\bm k,\alpha}^{\dagger} \;\bm \tau_{\alpha\beta}^{+\theta/2} \cdot (\bm k - \frac{\Delta\bm K}{2}) \; c_{2,\bm k,\beta}]\\
+ \,t_{\perp} \sum_{\bm k,\bm G,\alpha,\beta} (&c_{1,\bm k + \bm G,\alpha}^{\dagger} \; T_{\alpha\beta}(\bm G) \; c_{2,\bm k,\beta} + H. c.)\notag\;,
\end{align}
where $( \bm \tau^{\gamma}_x,\bm \tau^{\gamma}_y ) =  e^{\ii\gamma\bm\tau_z/2} ( \bm \tau_x, \bm \tau_y ) e^{-\ii\gamma\bm\tau_z/2}  $, $\bm \tau_{x,y,z}$ being Pauli matrices. The separation between twisted cones is $\Delta \bm K = 2 |\bm K| \sin(\theta/2)  \left[ 0,1\right]$  with $\bm K = \tfrac{4 \pi}{3 a_g} \left[ 1,0\right] $. Interlayer hopping is restricted to wavevectors $\bm G = \{\bm 0,-\bm G_1,-\bm G_1-\bm G_2\} $ with $\bm G_1 =  |\Delta \bm K| \left[ \tfrac{\sqrt{3}}{2},\tfrac{3}{2}\right]$, $\bm G_2 =  |\Delta \bm K| \left[ -\sqrt{3},0\right]$, and
\begin{equation}\label{Hopping}
\begin{split}
T(\bm 0) &= \begin{bmatrix} 1 & 1\\1 & 1\end{bmatrix},  
\\ T(-\bm G_1) &= T^{*}(-\bm G_1-\bm G_2)=\begin{bmatrix} 
e^{\ii 2 \pi / 3} & 1\\ e^{-\ii 2 \pi / 3}& e^{\ii 2 \pi / 3}\end{bmatrix}
.\end{split}
\end{equation}
The Hamiltonian is described by two parameters, $v_F $ and $ t_{\perp}$. The Fermi velocity is connected with the tight-binding Hamiltonian  by the relations $\hbar v_F= \tfrac{\sqrt{3}}{2} |t| a_g $, whereas $t_{\perp} $ can be obtained from the Fourier transform of the tight-binding interlayer Hamiltonian as described in appendix \ref{inter}. Calculations correspond to the choice $t_{\perp}=0.12\,\text{eV} $. 

Parallel currents are restricted to graphene layers, where they become the pseudospin operators. They are denoted $\bm j^{(1,2)} $, as in the tight-binding model. For instance, the $\bm q=0$, $x$ component of the current density for layer $(1)$ is given by 
\begin{equation}\label{current}
\hat{\bm x}\cdot \bm j^{(1)} = \frac{e \, v_F}{S}\sum_{\bm k,\alpha,\beta} c_{1,\bm k,\alpha}^{\dagger} \;\tau^{x}_{\alpha\beta} \; c_{1,\bm k,\beta} 
,\end{equation}
with Pauli matrix  $\tau^{x}= \bigl(\begin{smallmatrix}0&1\\1&0\end{smallmatrix}\bigr)$, and straightforward generalization to the remaining cases.

Linear response to the perturbing fields, $\bm A^{(1,2)} $,  proceeds as usual. Diamagnetic currents are nominally absent, though the treatment of  the ultraviolet cut-off requires some care if one is to extract the Drude weight  from the usual optical conductivity\cite{Stauber13,Moon13}. The fact that only two currents and two perturbing fields are present implies $4 \times 4$ response matrices, for which  we keep the same tight-binding notation. For instance, the Drude matrix in the continuum model has the block structure
\begin{equation}\label{Drudecont}
\bm D = 
	\left[
	\begin{array}{cc|cc}
	 D_0 & 0      &  D_1       & D_{xy}  \\
	  0     & D_0 & -D_{xy}  & D_{1}   \\
	\hline
	  D_1  & -D_{xy}   &  D_0 & 0       \\
	 D_{xy}  & D_{1} &   0     & D_0   \\
	\end{array}
	\right]
.\end{equation}

 Except for the obvious reduction of Drude terms, the entire discussion of section \ref{physical} applies to the continuum case. Therefore, Eqs. \ref{constit}  still applies, but with Drude terms given by
\begin{align}
 & D_T  = 2 (D_0+D_1)  \label{DTcont} \\
 & D_{chir}  = D_{xy}  \label{Dchircont} \\
 & D_{mag}  = D_1 - D_0 \label{Dmagcont}
,\end{align}
in the continuum model.

 As for the Drude case, the equilibrium response in the continuum model becomes the $4\times 4$ matrix
\begin{equation}\label{chieqcont}
\bm \chi_{eq} = 
	\left[
	\begin{array}{cc|cc}
	 \tilde \chi_0 & 0      &  \tilde \chi_1     & \tilde \chi_{xy} \\
	  0     & \tilde \chi_0 & -\tilde \chi_{xy}  & \tilde \chi_{1}  \\
	\hline
	  \tilde \chi_1  & -\tilde \chi_{xy}   &  \tilde \chi_0 & 0     \\
	 \tilde \chi_{xy}  & \tilde \chi_{1} &   0     & \tilde \chi_0   \\
	\end{array}
	\right]
,\end{equation}
and the corresponding gauge invariance requirements are
\begin{align}
\tilde \chi_0 + \tilde \chi_1 & = 0 \\
\tilde \chi_{xy}  & = 0 \label{gaugecont}
.\end{align}

\section{RESULTS}\label{RS}

\subsection{Drude matrix}

The comparison between the tight-binding and the continuum model results is presented in this section as a function of chemical potential. We will restrict our attention to the region around zero doping. Needless to say, the validity (and its limits) of the continuum  description of single-layer graphene is taken for granted. What is at the stake here is, therefore, mainly an assessment of the  approximate description of the interlayer Hamiltonian in the continuum model, mostly for linear response. 

\begin{figure}
\includegraphics[width=\columnwidth]{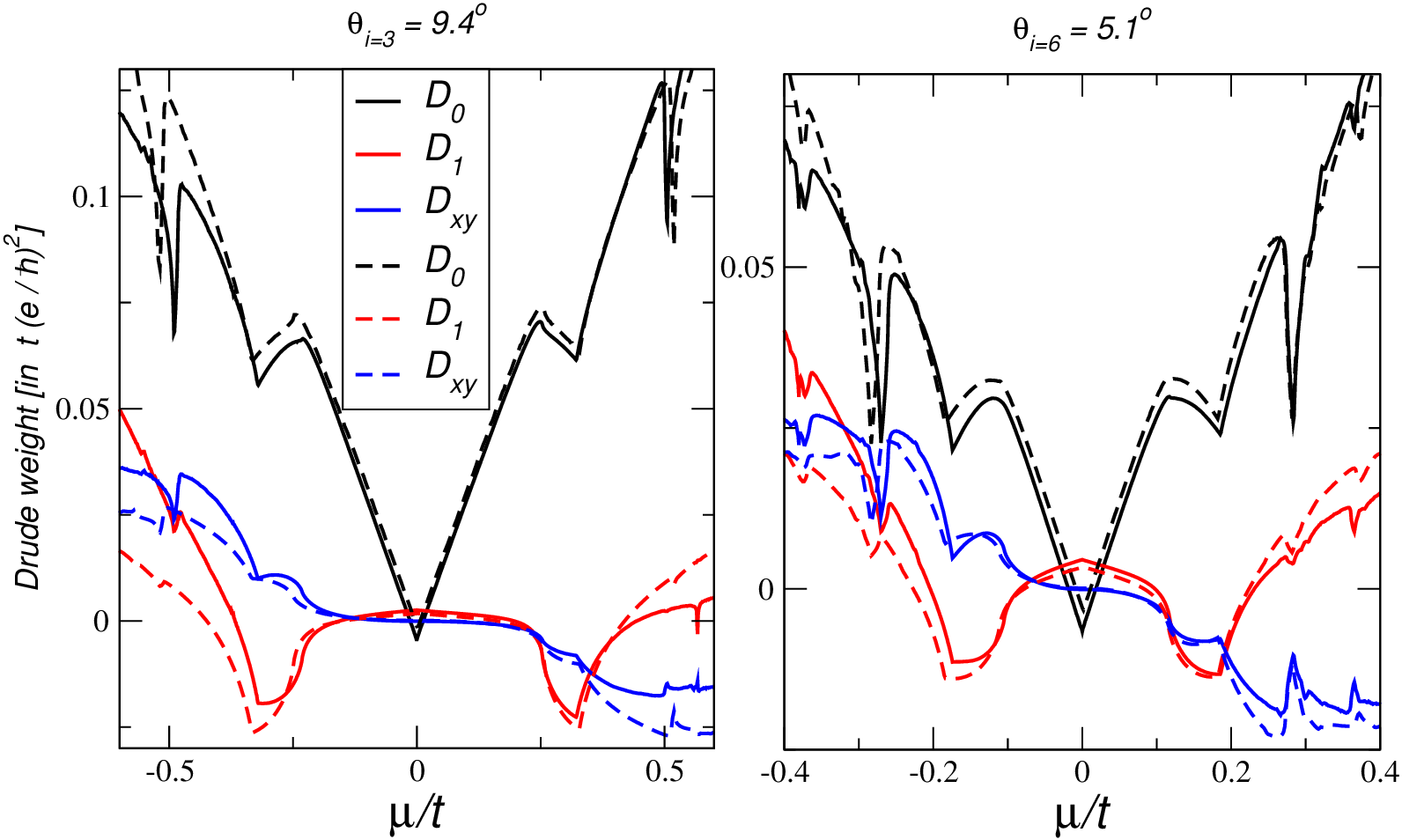}
\caption{(color online):  $D_0$ (black), $D_1$ (red) and $D_{xy}$ (blue)  entries of the  Drude matrix for the  tight-binding  (solid lines) and continuum (dashed lines) models as functions of the chemical potential. Left panel: twist angle $\theta_{i=3} = 9.4^o$. Right panel: twist angle $\theta_{i=6} = 5.1^o$.
\label{Fig1}}
\end{figure}

The simplest comparison corresponds to the common Drude entries of both models, namely, $D_0, D_1$, and  $D_{xy}$. They are shown in Fig. \ref{Fig1} as a function of chemical potential for two twist angles. Though  quantitative differences are visible, mainly a systematic greater electron-hole asymmetry in the tight-binding model, the overall behavior is very similar in both models. All the qualitative relevant features reported by us before for the continuum model, are present in the tight-binding calculation. For instance  the very existence of a chiral term $D_{xy} $, and its Hall-like dependence on carrier sign is preserved in the tight-binding results. The same applies to the term  $D_1 $: its dependence upon doping and its offset above $D_0 $ at zero doping, related later to paramagnetism,  are also  systematic features of the tight-binding results.    

\begin{figure}
\includegraphics[width=\columnwidth]{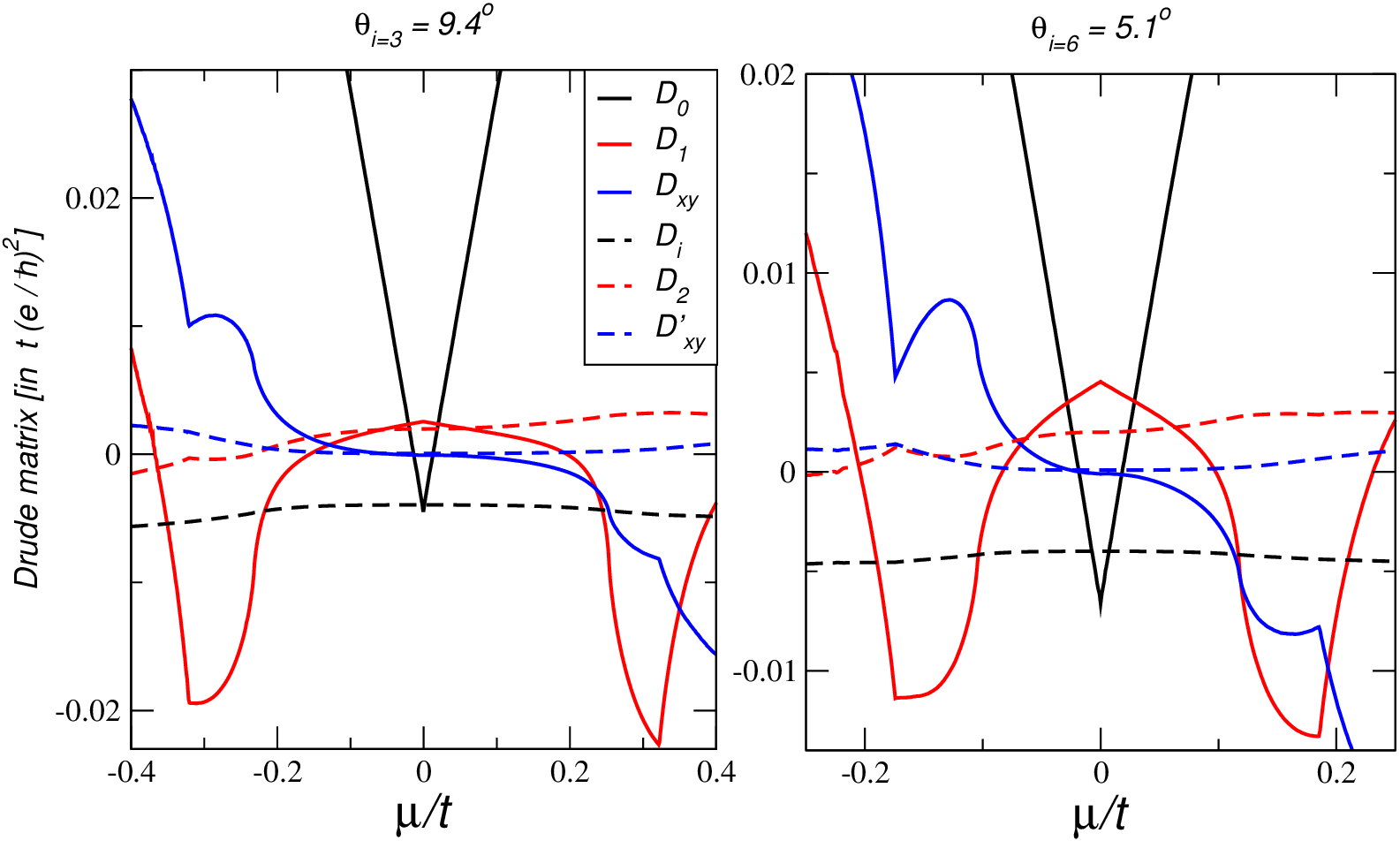}
\caption{(color online): All entries of the tight-binding Drude matrix as functions of the chemical potential: $D_0$ (solid black), $D_1$ (solid red), $D_{xy}$ (solid blue), $D_{i}$ (dashed black), $D_{2}$ (dashed red), and $D'_{xy}$ (dashed blue). Left panel: twist angle $\theta_{i=3} = 9.4^o$. Right panel: twist angle $\theta_{i=6} = 5.1^o$.
\label{Fig2}}
\end{figure}

The remaining entries of the tight-binding Drude matrix, $D_i, D_2, D'_{xy}$, are connected with the interlayer parallel current, neglected in the continuum. They are presented  in Fig. \ref{Fig2}, where they are compared with $D_0, D_1, D_{xy} $. They are generally smaller and featureless in that range, though $D_i $ can become sizable near zero-doping.

\begin{figure}
\includegraphics[width=\columnwidth]{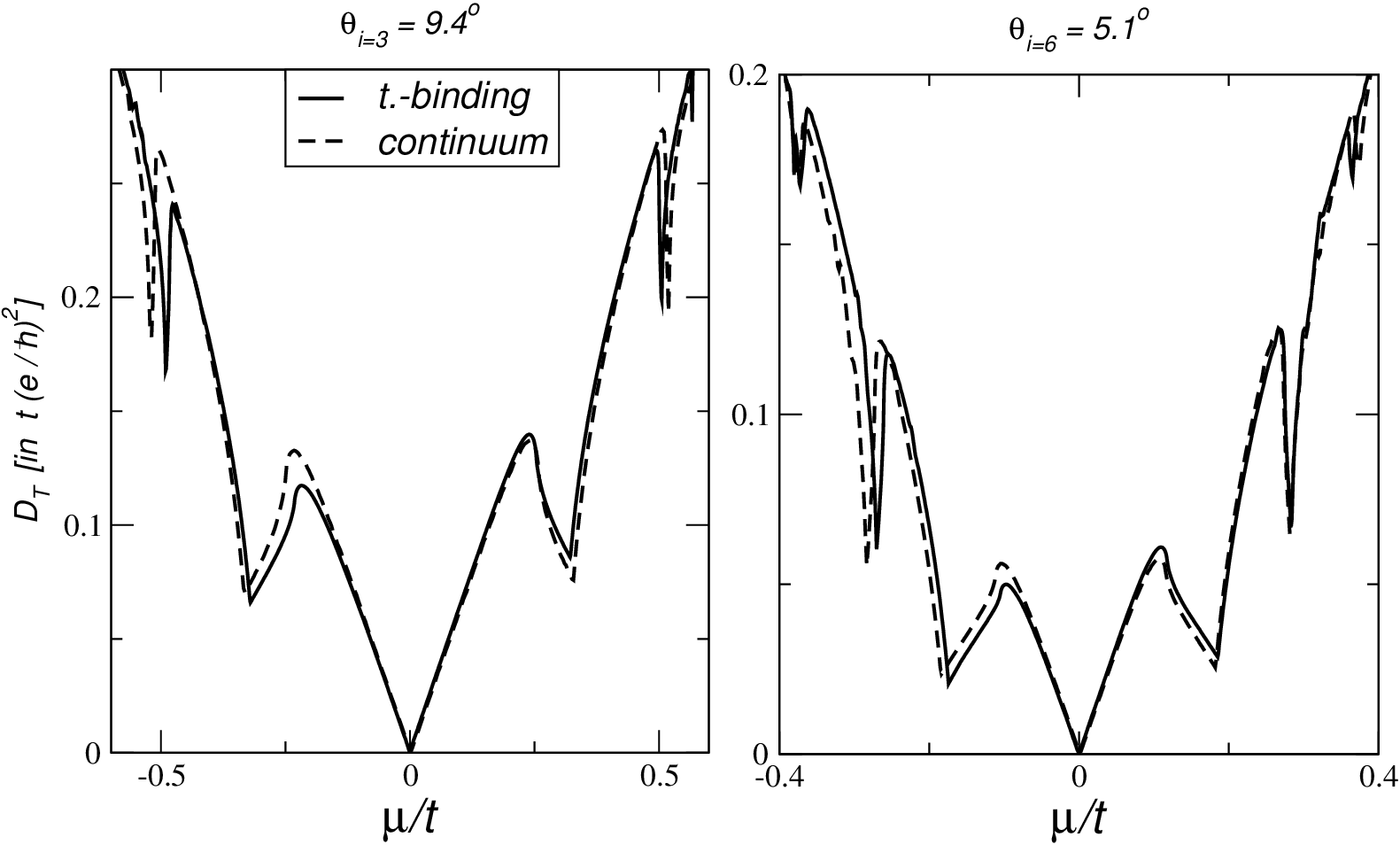}
\caption{ Total Drude weight as function of the chemical potential  for the tight-binding  (solid line) and continuum models (dashed line). Left panel: twist angle $\theta_{i=3} = 9.4^o$. Right panel: twist angle $\theta_{i=6} = 5.1^o$.
\label{Fig3}}
\end{figure}

 Perhaps a more sensible comparison from a physical standpoint  is afforded by the parameters $D_T, \, D_{chi}$, and $ D_{mag} $. They describe the physical response in exactly the same way for both models, Eqs. \ref{constit}. 
The total Drude weight, $D_T $, first considered in Ref. \cite{Stauber13}, is presented in Fig. \ref{Fig3} for both models. Notice that $D_T$ describes the total current accelerated by an electric field, and could have been obtained from the mass tensor of the band structure, as shown in the appendix \ref{Lresponse}. The agreement between both models is remarkable. 

\begin{figure}
\includegraphics[width=\columnwidth]{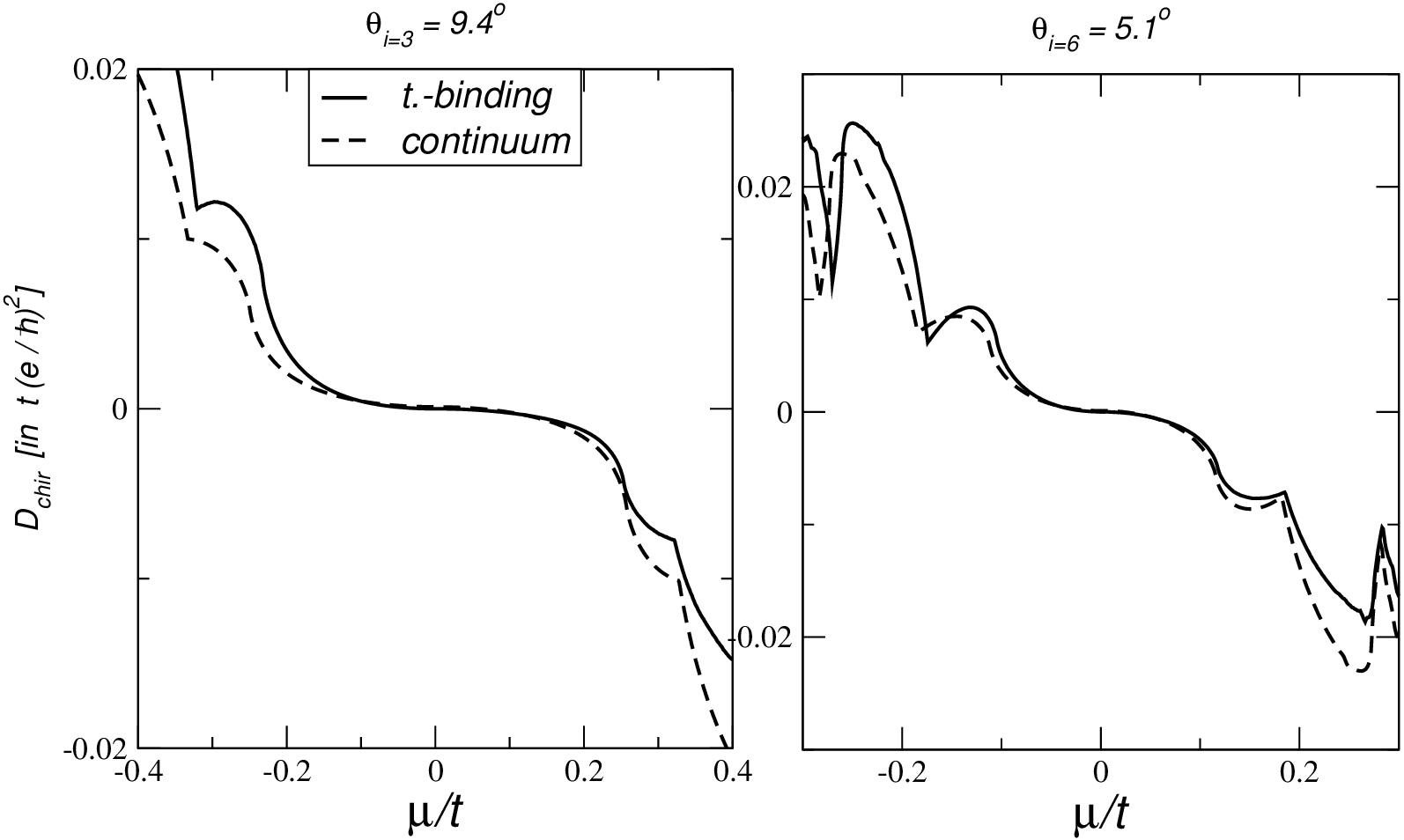}
\caption{ Chiral Drude component as function of the chemical potential  for the tight-binding  (solid line) and continuum models (dashed line). Left panel: twist angle $\theta_{i=3} = 9.4^o$. Right panel: twist angle $\theta_{i=6} = 5.1^o$.
\label{Fig4}}
\end{figure}

The chiral contributions, Eqs. \ref{Dchir}  and  \ref{Dchircont}, are compared in Fig. \ref{Fig4}. As already mentioned, the qualitative behavior is very similar. Therefore, the main physical significance of this chiral term, namely, the parallel magnetic moment accompanying the longitudinal currents of intrinsic plasmonic excitations, Eq. \ref{constraint}, seems to be a robust feature of the system.

\begin{figure}
\includegraphics[width=\columnwidth]{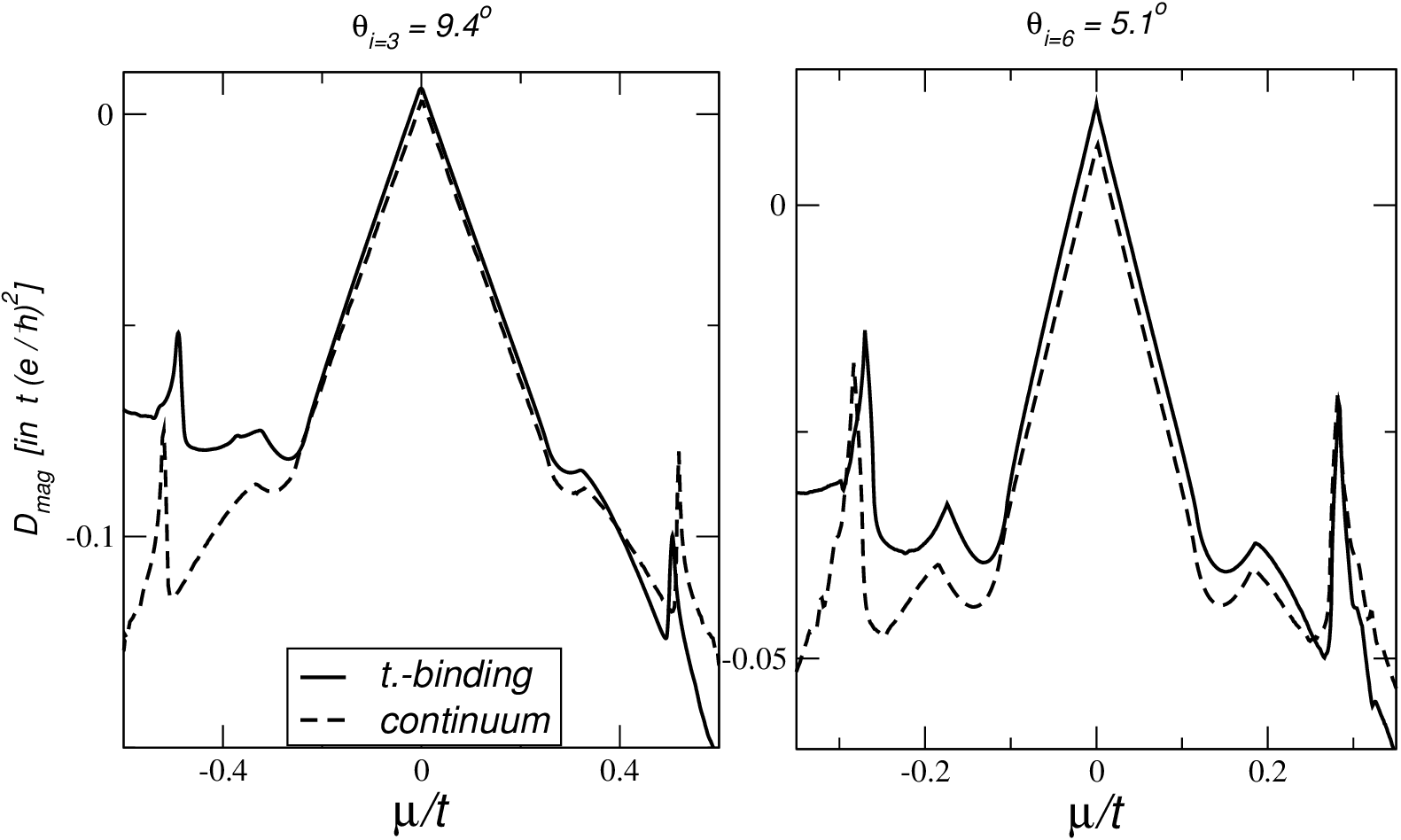}
\caption{ Magnetic Drude component as function of the chemical potential  for the tight-binding  (solid line) and continuum models (dashed line). Left panel: twist angle $\theta_{i=3} = 9.4^o$. Right panel: twist angle $\theta_{i=6} = 5.1^o$.
\label{Fig5}}
\end{figure}

Finally, the comparison for the parameter $D_{mag}$ is shown in Fig. \ref{Fig5}.  Owing to its definition, $ D_{mag} \propto(D_1-D_0) $, it can be interpreted as the Drude weight for accelerating opposite currents in each layer, or counterflow. Accounting for the (magnetic) sign convention of Eq. \ref{constit0}, the mostly  negative $D_{mag} $ of Fig. \ref{Fig5} implies that the current in each layer is accelerated by their respective electric field in the expected {\em correct way}. But, as noted in our previous work for the continuum model, $D_{mag}  $ starts off positive and remains so  in a finite range around the neutrality point, a  feature  also confirmed here in the tight-binding calculation. This implies that, within that range, the electric field is accelerating currents in the apparently {\em wrong way} and that, even at the neutrality point, there are couterflow currents. This puzzling picture is made more conventional in the magnetic language of Eqs. \ref{constit}, where it could also be seen as  the emergence of a magnetic moment upon the slow application of a magnetic field, for which the sign of the response need not be  prejudiced,  and free carriers need not be present, as neutral graphene shows. Both models  give a positive sign at the neutrality point for the twist angles here considered, implying paramagnetism. Indeed, we will later see that in-plane orbital paramagnetism is also the equilibrium susceptibility for a rather wide doping window.

\begin{figure}
\includegraphics[width=\columnwidth]{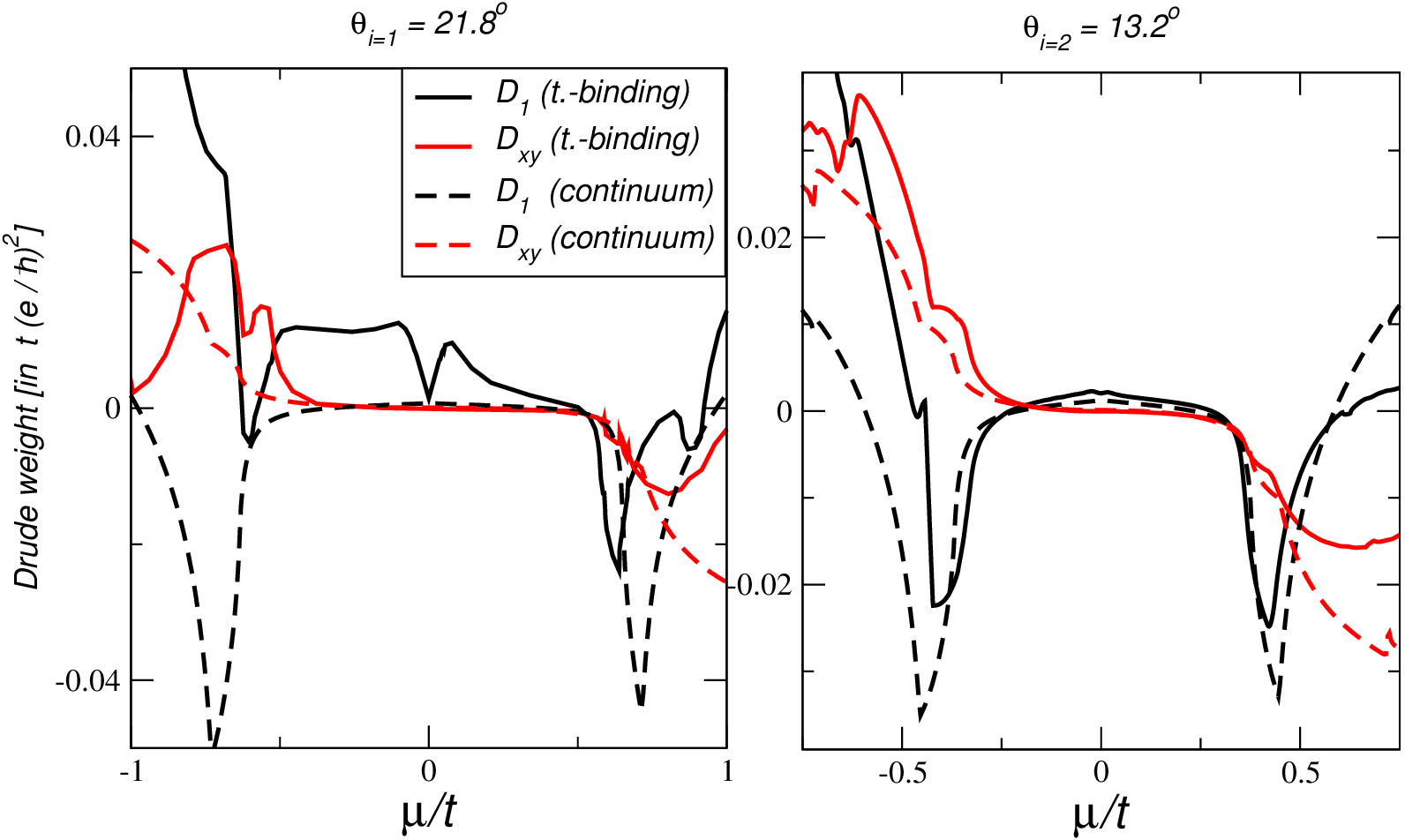}
\caption{(color online):  $D_0$ (black), $D_1$ (red) and $D_{xy}$ (blue)  entries of the  Drude matrix for the  tight-binding  (solid lines) and continuum (dashed lines) models as functions of the chemical potential. Left panel: twist angle $\theta_{i=1} = 21.8^o$. Right panel: twist angle $\theta_{i=2} = 13.2^o$.
\label{Fig6}}
\end{figure}

From the above analysis, it is clear that tight-binding and continuum models agree on the basic aspects. It is true, however, that the tight-binding numerical effort  limits the  accessible angles. As argued in ref.\cite{Lopes12}, though, the very nature of the continuum model suggests its   becoming increasingly better for smaller angles. From this perspective, the  comparison should degrade for  larger commensurate angles. This is shown in  Fig. \ref{Fig6}, where the lowest commensurate structures are shown, $\theta_{i=1}=21.8^o $ in the left and 
$\theta_{i=2}=13.2^o $ in the right. For such large angles the interlayer coupling is very small, and only  the interlayer dominated entries $D_1$ and $D_{xy}$ are shown.  For $\theta_{i=1}=21.8^o $, significant goodwill is required to  discover  similarities between tight-binding and continuum. But for $\theta_{i=2}=13.2^o  $, the comparison dramatically improves, with all the salient qualitative features considered above clearly present. Looking at Fig. \ref{Fig1}, one could say that $\theta_{i=3}=9.4^o  $ marks the beginning of quantitative agreement.


\subsection{Equilibrium response. Parallel magnetic field}

Here we consider the true equilibrium response  and explore the fate of expressions like those of Eqs. \ref{adiabaticj} and \ref{adiabaticm}. 
A parallel magnetic field can be introduced by the following choice of  perturbing vector potential
\begin{equation}\label{bfield}
\bm A^{(1)}=\frac{a}{2} \hat{\bm z} \times \bm B_{\parallel}, \;\; \bm A^{(2)}=-\bm A^{(1)},\;\;  \bm A^{(inter)}=0
.\end{equation}
It is worth mentioning that, although we will use the linear response formalism  outlined in section \ref{equilresp}, one could alternatively calculate currents directly from the  ground state averages of the perturbed Hamiltonian. The reason being that no computational penalty arises in the Hamiltonian perturbed  by the vector potential of Eq. \ref{bfield}, as it retains the original translational symmetry. In fact, we have often used this second route as an additional consistency check.

We first consider  the equilibrium version of Eq. \ref{adiabaticm}, 
\begin{equation}\label{mequil}
  \bm m_{\parallel}  =   \frac{a^2}{2} \tilde \chi_{mag} \;  \bm B_{\parallel} 
,\end{equation}
where
\begin{equation}\label{chimag}
  \tilde \chi_{mag} =  \tilde \chi_1 - \tilde \chi_0 
.\end{equation}
\begin{figure}
\includegraphics[width=\columnwidth]{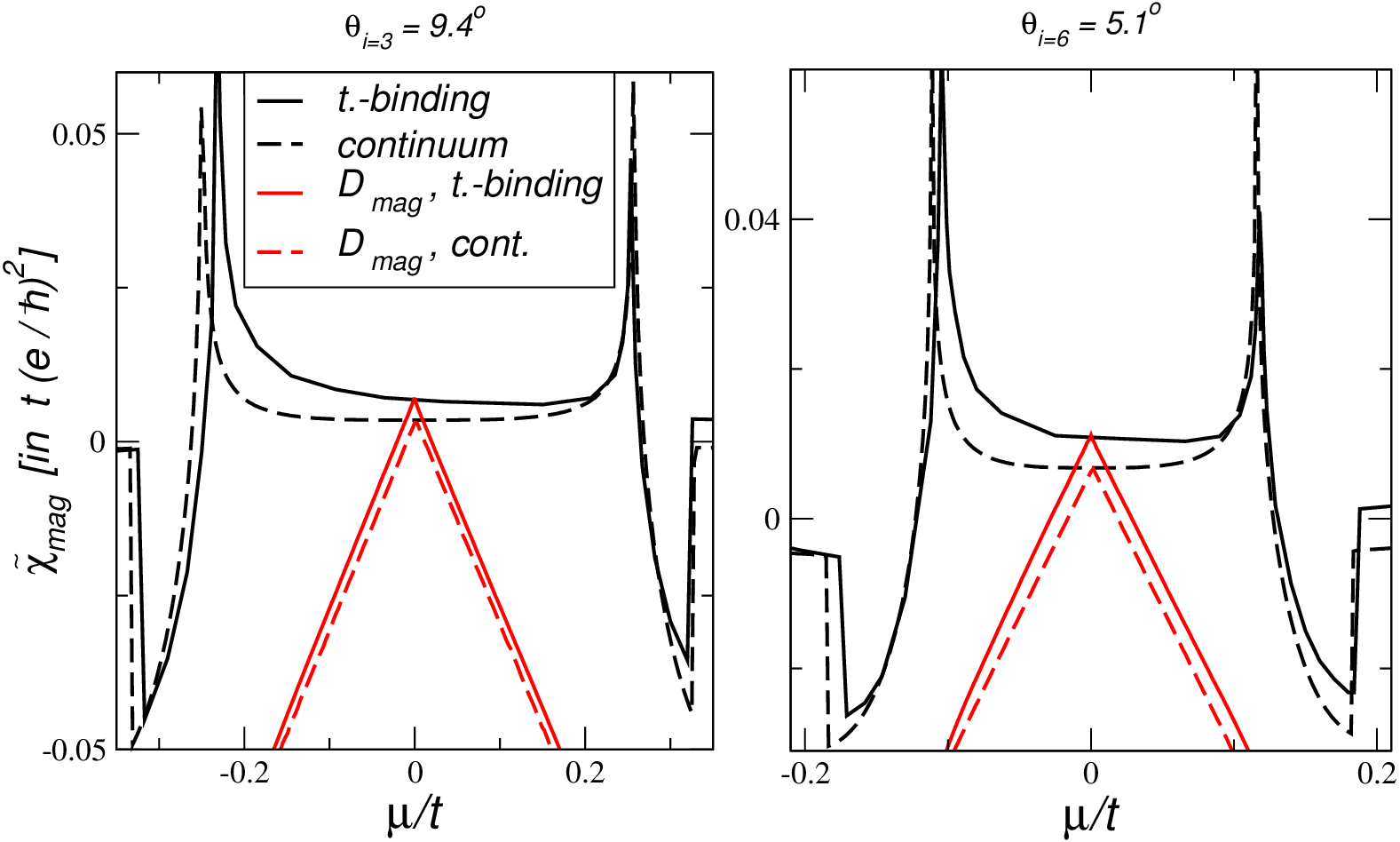}
\caption{(color online): Equilibrium magnetic response to a parallel magnetic field for the tight-binding  (solid black line) and  the  continuum (dashed black line) models as functions of the chemical potential. The corresponding Drude responses of Fig. \ref{Fig5} are included for comparison (red lines). Left panel: twist angle $\theta_{i=3} = 9.4^o$. Right panel: twist angle $\theta_{i=6} = 5.1^o$.
\label{Fig7}}
\end{figure}
In Fig. \ref{Fig7},  we plot the equilibrium susceptibility as a function of chemical potential. Albeit with some quantitative differences, both tight-binding and continuum cases exhibit similar behavior. There is a  positive response  in an extended plateau around the neutrality point, roughly covering the entire region between  the energies corresponding to the intersecting Dirac cones. Therefore, the equilibrium magnetic response in that area corresponds to (orbital) paramagnetism. 
The gate dependence of the magnetic response of Fig. \ref{Fig5} is  
strikingly similar to gate dependence of the lattice contribution of  
the out-of-plane magnetic susceptibility of single layer  
graphene\cite{Gomez11} and related systems.\cite{Raoux15,Gutierrez16}  
This points to some sort of universality in the orbital response of  
layered materials which seems to be independent of the field direction  
and would deserve further investigation.

For comparison, the Drude response of Fig. \ref{Fig7} is also plotted, showing that Drude and equilibrium response coincide at the neutrality point, where the Fermi surface correction vanishes, as expected. 
Aspreviously reported,\cite{Stauber18} this orbital paramagnetism can be quite substantial if compared to other sources of orbital magnetic response, in the vicinity of the magic twist angle.\cite{Erratum} Furthermore, the vanishing of the density of states and  Pauli spin paramagnetism, makes this orbital paramagnetism the dominant  response around the neutrality point.  

We now inquire about the possible existence of an equilibrium counterpart to Eq. \ref{adiabaticj},
\begin{equation}
\bm j_T  =    - a \tilde \chi_{chir}  \;  \bm B_{\parallel}  \label{adiabaticjeq} 
,\end{equation}
where  now
\begin{equation}\label{chiraleq}
\tilde \chi_{chir} =\begin{cases}
  \tilde \chi_{xy} + \tilde \chi'_{xy} & \text{tight-binding},\\
   \tilde \chi_{xy}& \text{continuum}.
\end{cases}
\end{equation}
Let us recall that both in the Drude and equilibrium cases, the emergence of a parallel current in response to a parallel magnetic field is allowed on time and (lack of) parity symmetry.  In spite of this, the gauge invariance relations Eqs. \ref{gaugetight} and \ref{gaugecont} make
\begin{equation}
\tilde \chi_{chir} =0
,\end{equation}
and, therefore, the total equilibrium current vanishes. 

\begin{figure}
\includegraphics[width=\columnwidth]{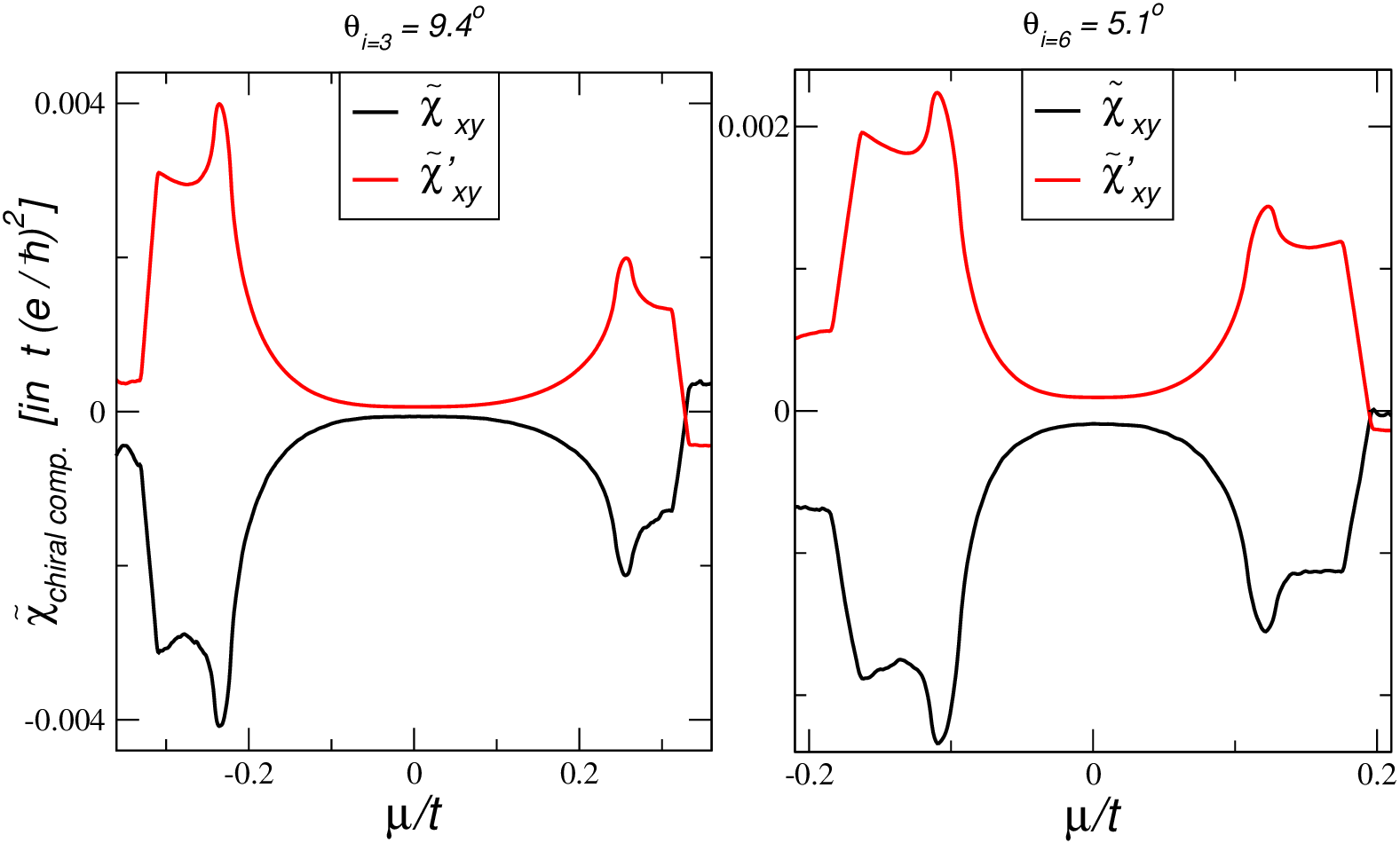}
\caption{(color online): Chiral components of the equilibrium  response, $\tilde \chi_{xy}$ and $\tilde \chi'_{xy} $, for the tight-binding calculation (see Eqs. \ref{adiabaticjeq} and \ref{chiraleq}). Left panel: twist angle $\theta_{i=3} = 9.4^o$. Right panel: twist angle $\theta_{i=6} = 5.1^o$.
\label{Fig8}}
\end{figure}

However, it is interesting to realize that the cancellation of $\tilde \chi_{chir}$  takes place with non-zero values of $ \tilde \chi_{xy}$ and $\tilde \chi'_{xy} $ in the tight-binding case, as shown in Fig. \ref{Fig8}. This means that, though globally zero, there is a current structure summarized as follows
\begin{equation}\label{currentstruct}
\bm j^{(1)} = \bm j^{(2)} = - \frac{1}{2} \bm j^{(inter)}
.\end{equation}
That is, the parallel current associated to the non-vertical nature of the interlayer bonds is non-zero, and opposite to that carried by the layers themselves. The current structure  illustrated in Fig. \ref{Fig8} is a consistent  feature of all our tight-binding calculations. Notice that, were the system parity invariant, each such current contribution would be forbidden. Therefore, this layered current response to a magnetic field is a remainder of the chiral nature of TBL. 
%

%\subsection{Drude $D_T$ and superfluid density? Results for magic angle?}

\subsection{Drude weight and superfluid density}

In view of the recent developments concerning superconductivity in TBG,\cite{Cao18b} it is worth closing this section by recalling that the BCS gap makes the difference between Drude and equilibrium responses disappear in the  superconducting ground state. Thus, for instance, $D_T$ would become the physically correct {\it equilibrium} response to an in-plane {\it transverse} vector potential, that is, the superfluid density\cite{Scalapino93} $D_S$. For the usual case of a superconducting gap much smaller than the bandwidth,  also applicable to superconducting TBG, the quantitative evaluation of the superfluid density at zero temperature could be carried out in the normal state. Therefore,  a normal state calculation of $D_T$ like that of Fig. \ref{Fig3}  close to the magic angle  could be immediately translated as   the  superfluid density of the superconducting ground state.

\section{SUMMARY}\label{SM}

We have presented a comprehensive study of the electromagnetic linear response of TBL, described by both a tight-binding model and its continuum limit. The study has been restricted to homogeneous horizontal fields, $ \bm q \to 0$, but otherwise unrestricted along the stacking direction. This non-locality along the $\hat{\bm z}$, which is a requirement to describe  optical activity at finite frequencies, has been here studied in the limit $\omega \to 0$, on the grounds that novel phenomena might be unearthed by the experimental possibility of addressing  layers individually. Our study has considered both the dynamical, Drude aspect ($\omega=0$ limit first)  and the equilibrium response  ($\bm q=0$ limit first).

As for the declared objective of assessing the validity of the continuum model, the conclusion is affirmative: all previously reported\cite{Stauber18} qualitative features on the continuum model are present in the tight-binding calculation.
In particular, the tight-binding calculation fully supports the existence of a peculiar magnetic or counterflow Drude component, $ D_{mag}=D_1-D_0$, finite even at the neutrality point and with  nominally wrong sign. The agreement also extends to the chiral Drude component, $D_{chir}$, implying that accelerated currents such as those of intrinsic plasmons are accompanied of a parallel magnetic moment, the basic signature of chirality. The calculation has been extended to cover the equilibrium response, where the agreement between tight-binding and continuum models also holds. The equilibrium response to a parallel magnetic field implies  orbital paramagnetism over a substantial doping range and the existence of a layered current structure as the last remnant of  chirality.

\subsection{ACKNOWLEDGMENTS}
Support from Spain's MINECO  Grants No. MDM-2014-0377, No. FIS2017-82260-P, and No. FIS2015-64886-C5-5-P is gratefully acknowledged. TL acknowledges support by the National Science Foundation NSF/EFRI grant (\#EFRI-1741660).

\appendix
\section{Interlayer Hamiltonian}\label{inter}
The tight-binding hopping parameter  between two $p_z $ orbitals in different layers is taken to be 
\begin{equation}
V(d) = c \left[ \left(\frac{a}{d}\right)^2 \, V_{pp\sigma}(d) + \left(\frac{\rho}{d}\right)^2 \, V_{pp\pi}(d)\right]
,\end{equation}
where $d = \sqrt{\rho^2 + a^2} $ is their distance, with in-plane component $\rho$ and   interlayer separation $a$. Adapting the treatment of ref.\cite{Tang96}, $ V_{pp\sigma}(d)$ and $ V_{pp\pi}(d)$  are assumed to depend on distance as
\begin{equation}
\begin{split}
 V_{pp\sigma}(d) &= \frac{\alpha_1}{d^{\alpha_2}} \exp (-\alpha_3 \, d^{\alpha_4})  \\
 V_{pp\pi}(d)    &=  \frac{\beta_1}{d^{\beta_2}} \exp (-\beta_3 \, d^{\beta_4})
,\end{split}
\end{equation}
with $ \alpha_1=11.7955,\,  \alpha_2=0.7620,\,\alpha_3=0.1624,\,\alpha_4=2.3509 $  , and $ \beta_1=-5.4860,\,  \beta_2=1.2785,\,\beta_3=0.1383,\,\beta_4=3.4490 $, in $\text{eV}$ and $\mathring{\text{A}} $ units. The interlayer distance has been taken as $a=3.5 \, \mathring{\text{A}}$, and the graphene lattice constant as $a_g=2.46 \, \mathring{\text{A}}$.
The  overall constant $c$ is adjusted so that the 2d Fourier transform
\begin{equation}\label{2dfourier}
\tilde{V}(\bm q) = \frac{1}{s_0} \int d^2 \rho  \; e^{-\ii \bm q \cdot \bm \rho}  \; V(\sqrt{\rho^2 + a^2})
,\end{equation}
evaluated at the Dirac K-point with $\bm K = \tfrac{4 \pi}{3 a_g}(1,0) $, gives $\tilde{V}(\bm K) = 0.12 \, \text{eV}$. $s_0$ is the graphene unit cell area. 
This interlayer scheme produces for the hopping integral between two vertically aligned orbitals the value $t_{A_1B_2}=0.49 \, \text{eV} $, very close to that used in previous tight-binding calculations\cite{Moon13}. 
%
%, around the often assumed\cite{Li10,Bistritzer11,Lopes07}  $10$ percent  of the intralayer hopping Hamiltonian.  
%

Notice that $\tilde{V}(\bm K)$ is the magnitude that appears  in the continuum model for the interlayer Hamiltonian, as shown in ref.\cite{Bistritzer11} . Therefore, the quantitative connection between the tight-binding model and the continuum model for the interlayer term is
\begin{equation}\label{connect}
t_{\perp} = \tilde{V}(\bm K) = 0.12 \, \text{eV}
.\end{equation}

With the choice of Eq. \ref{connect}, one has the ratio $\tfrac{t_{\perp}}{|t|} \sim 0.4 $, as in previous continuum model calculations.\cite{Lopes07,Stauber13,Stauber18}. This choice also produces for the first magic angle\cite{Bistritzer11} the value $\theta\sim\theta_{i=31}=1.05^o$.

\section{Tight-binding Linear Response}{\label{Lresponse}

Any tight-binding Hamiltonian can accommodate the presence of an electromagnetic field, given by the vector potential $\bm A $, by the following replacement for each elementary hopping term
\begin{equation}\label{Peierls}
t_{ij} \, c^{\dagger}_i \, c_j \rightarrow  t_{ij} e^{-\ii \tfrac{e}{\hbar}\bm A \cdot \bm r_{ij}} \, c^{\dagger}_i \, c_j 
,\end{equation}
with $\bm r_{ij} = \bm r_j - \bm r_i $, where $\bm r_{j(i)} $ are the orbital positions, and $\bm A $, the average field along the bond. Currents operators are then obtained for each bond from the functional derivative $\bm j = -\tfrac{ \partial \mathcal{H}}{\partial \bm A} $. This leads  to the following expression for the current operator associated with an elementary hopping term:
\begin{equation}\label{bondcurrent}
\bm j_{ij} = \bm j_{p,ij} + \bm j_{d,ij}
,\end{equation}
where the first term defines the paramagnetic current operator, given by  
\begin{equation}\label{bondp}
\bm j_{p,ij} = \ii \frac{e}{\hbar} \bm  r_{ij}\; t_{ij} c^{\dagger}_i \, c_j 
,\end{equation}
 and the second is the diamagnetic one, given to linear order by
\begin{equation}\label{bondd}
\bm j_{d,ij} =   \frac{e^2}{\hbar^2} \; t_{ij} c^{\dagger}_i \, c_j \bm  r_{ij} \bm  r_{ij} \cdot \bm A
.\end{equation}

\subsection{$\bm q=0$ response. Drude limit}
Summing Eq. \ref{bondp} for all hopping terms, then the $\bm q=0 $ Fourier component of the parallel, paramagnetic current operator can be decomposed as 
\begin{equation}\label{jp}
\begin{split}
\bm j_p^{(1)} =& \frac{e}{S} \sum_{\substack{\bm k\\ n\in 1, m\in 1}}  \bm v_{nm}(\bm k) c^{\dagger}_{\bm k, n} c_{\bm k, m} \\
\bm j_p^{(2)} =& \frac{e}{S} \sum_{\substack{\bm k\\ n\in 2, m\in 2}}  \bm v_{nm}(\bm k) c^{\dagger}_{\bm k, n} c_{\bm k, m} \\
\bm j_p^{(inter)} =& \frac{e}{S} \; [\sum_{\substack{\bm k\\ n\in 1, m\in 2}} \bm v_{nm}(\bm k) c^{\dagger}_{\bm k, n} c_{\bm k, m} \\
&+ \sum_{\substack{\bm k\\ n\in 2, m\in 1}}  \bm v_{nm}(\bm k) c^{\dagger}_{\bm k, n} c_{\bm k, m} ]
.\end{split}
\end{equation}
 $\bm j_p^{(1,2)} $ correspond to the intralayer currents whereas  $\bm j_p^{(inter)} $ describes the parallel current carried by the (oblique) interlayer tight-binding {\em bonds}. $c^{\dagger}_{\bm k, n} (c_{\bm k, n} )$ are fermion operators for the Bloch state with orbital index $n$. The velocity matrix is given by
\begin{equation}
\bm v_{nm}(\bm k) = \hbar^{-1} \grad_{\bm k} h_{nm}(\bm k)
,\end{equation}
where  $h_{nm}(\bm k) = \mel{\bm k,n}{\mathcal{H}_0}{\bm k,m}$ is the Bloch matrix in orbital indices,  and  $\ket{\bm k,n} $, the Bloch state for supercell orbital index $n$.

The response tensor $\bm \chi_p$ for $\bm q=0 $ enjoys all the symmetries of the problem, namely, time-reversal for $\mathcal{H}_0$, rotational invariance around the $\hat{\bm z}$ axis, and $\pi$-rotation invariance around any in-plane axis in the mid-point between layers. As a consequence, non-zero entries are those of Eq. \ref{chipbis}. Linear response dictates their generic form to be as follows
\begin{equation}\label{Kubo}
\begin{split}
\chi_p(\omega) = 
  S \sum_{\bm k, n, m} &
\langle m,\bm k|A|n,\bm k\rangle \langle n,\bm k|B|m,\bm k\rangle \times \\
& \frac{n_F(\epsilon_{m,\bm k}) - n_F(\epsilon_{n,\bm k})}{\hbar \omega_+  - \epsilon_{n,\bm k} + \epsilon_{m,\bm k}}
,\end{split}
\end{equation}
where $\omega_+ = \omega  +\ii 0^+$, and the  states $|m, \bm k\rangle$ are Bloch eigenstates\cite{Notation} of $\mathcal{H}_0$
with  {\em band} index $m$ and eigenenergy are $\epsilon_{n,\bm k} $, and  $n_F$ is the Fermi function.
The operator correspondences for each entry are:
\begin{align}\label{oper}
%\begin{split}
 \chi_0: \;\;\; & A=\hat{\bm x}\cdot \bm j^{(1)}_p \, \text{and} \,  B=\hat{\bm x}\cdot \bm j^{(1)}_p \notag \\
 \chi_i: \;\;\; & A=\hat{\bm x}\cdot \bm j^{(inter)}_p \, \text{and} \,  B=\hat{\bm x}\cdot \bm j^{(inter)}_p \notag\\
 \chi_1: \;\;\; & A=\hat{\bm x}\cdot \bm j^{(1)}_p \, \text{and} \,  B=\hat{\bm x}\cdot \bm j^{(2)}_p \notag\\
 \chi_2: \;\;\; & A=\hat{\bm x}\cdot \bm j^{(1)}_p \, \text{and} \, B=\hat{\bm x}\cdot \bm j^{(inter)}_p \\
 \chi_{xy}: \;\;\; & A=\hat{\bm x}\cdot \bm j^{(1)}_p \, \text{and} \,  B=\hat{\bm y}\cdot \bm j^{(2)}_p \notag\\
 \chi'_{xy}: \;\;\; & A=\hat{\bm x}\cdot \bm j^{(1)}_p \, \text{and} \, B=\hat{\bm y}\cdot \bm j^{(inter)}_p \notag
%\end{split}
,\end{align}
where $ \hat{\bm x}$ and $  \hat{\bm y}  $ are in-plane orthogonal unit vectors. Furthermore,  the {\em chiral} entries $ \chi_{xy}$ and $\chi'_{xy} $ are odd functions of the twist angle $\theta_i $, whereas the rest are even functions. 

 The non-zero entries of the $\bm q=0$ diamagnetic response, Eq. \ref{chidbis}, are given by 
\begin{equation}\label{jd}
\begin{split}
\chi_{d0} =& \frac{1}{S} \; \frac{e^2}{\hbar^2} \sum_{\substack{\bm k\\ n\in 1, m\in 1}}  [\partial^2_{k_x} h_{nm}(\bm k)] \;  \langle c^{\dagger}_{\bm k, n} c_{\bm k, m} \rangle \\
\chi_{di} =& \frac{1}{S} \;  \frac{e^2}{\hbar^2} \;  \{\sum_{\substack{\bm k\\ n\in 1, m\in 2}} [\partial^2_{k_x} h_{nm}(\bm k)] \; \langle c^{\dagger}_{\bm k, n} c_{\bm k, m}\rangle \\
&+  \sum_{\substack{\bm k\\ n\in 2, m\in 1}}   [\partial^2_{k_x} h_{nm}(\bm k)] \; \langle c^{\dagger}_{\bm k, n} c_{\bm k, m}\rangle  \}
,\end{split}
\end{equation}
where $\langle \, \rangle$ imply equilibrium average for $\mathcal{H}_0$.  Both $\chi_{d0}$ and $\chi_{di} $ are even function of the twist angle. Notice that the diamagnetic response does not depend on $\omega$. Therefore, The Drude  limit of Eq. \ref{Drude} is given 
explicitly by 
\begin{equation}%\label{Drude}
\bm D = \lim_{\omega \to 0} \bm \chi_p(\omega) + \bm \chi_{d}
.\end{equation}
 The total Drude weight of Eqs. \ref{DT} and \ref{DTcont} can also be obtained from the bands by the familiar expression
\begin{equation}\label{DTbands}
D_T = \frac{1}{S} \; \frac{e^2}{\hbar^2} \sum_{\bm k, n}  [\partial^2_{k_x} \epsilon_{n,\bm k}] \; n_F(\epsilon_{n,\bm k})
.\end{equation}

}

\subsection{Equilibrium response}\label{appequilibrium}
The $\omega \to 0$ limit of the Drude matrix corresponds to an adiabatic application of fields, and needs not coincide with the equilibrium response. In general, one has
\begin{equation}
\bm j(\bm q,\omega) = - \bm \chi(\bm q,\omega)  \bm A(\bm q,\omega) 
,\end{equation}
and the equilibrium response corresponds to
\begin{equation}
\bm \chi_{eq} = \lim_{\bm q \to 0} \lim_{\omega \to 0} \bm \chi(\bm q,\omega)  
,\end{equation}
whereas the Drude matrix is
\begin{equation}
\bm D =  \lim_{\omega \to 0} \lim_{\bm q \to 0}\bm \chi(\bm q,\omega)  
,\end{equation}
and the order of limits matters in the paramagnetic current response for gapless systems. Fortunately, the difference  is a Fermi surface term that comes from the $n=m$, intraband contribution in Eq. \ref{Kubo}. It can be obtained from the relation
\begin{equation}%\label{Kubo}
\begin{split}
 &\lim_{\bm q \to 0} \lim_{\omega \to 0} \frac{n_F(\epsilon_{n,\bm k - \bm q /2}) - n_F(\epsilon_{n,\bm k + \bm q /2})}{\hbar \omega - \epsilon_{n,\bm k - \bm q /2} + \epsilon_{n,\bm k + \bm q /2}} = \\
&\lim_{\omega \to 0} \lim_{\bm q \to 0} \frac{n_F(\epsilon_{n,\bm k - \bm q /2}) - n_F(\epsilon_{n,\bm k + \bm q /2})}{\hbar \omega - \epsilon_{n,\bm k - \bm q /2} + \epsilon_{n,\bm k + \bm q /2}} -
 \delta(\epsilon_{n,\bm k}-\mu) = \\
 & - \delta(\epsilon_{n,\bm k}-\mu)
,\end{split}
\end{equation}
where zero temperature has being assumed for simplicity.

Therefore,  the equilibrium entries of Eq. \ref{chieq} can be obtained from the Drude ones as
\begin{align}\label{chifermi}
%\begin{split}
 \tilde \chi_0 &= D_0 + \chi^F_0 \notag \\
 \tilde \chi_i &= D_i + \chi^F_i      \notag\\
 \tilde \chi_1 &=  D_1 + \chi^F_1    \notag\\
 \tilde \chi_2 &=  D_2 + \chi^F_2    \\
 \tilde \chi_{xy} &= D_{xy} + \chi^F_{xy} \notag\\
 \tilde \chi'_{xy} &= D'_{xy} +  \chi'^F_{xy}  \notag
%\end{split}
,\end{align}
with Fermi surface contributions given generically by
\begin{equation}\label{fermicorr}
\chi^F_{\alpha} = 
  S \sum_{\bm k, n} 
\langle n,\bm k|A|n,\bm k\rangle \langle n,\bm k|B|n,\bm k\rangle \, \, \delta(\epsilon_{n,\bm k}-\mu) 
,\end{equation}
where the operator correspondences for each entry $\alpha $ is as in  Eq. \ref{oper}.

The equilibrium response in the continuum model, not considered in our previous Ref.\cite{Stauber18} , is as in Eqs.  \ref{chifermi} and \ref{fermicorr}, with the obvious changes in current operators and number of terms.

\bibliography{paper.bib}

\begin{thebibliography}{48}
\bibitem [{Err()}]{Erratum}%
  \BibitemOpen
  \href@noop {} {}\bibinfo {note} {We note that the scale of the left graph of
  Fig. 1 of the SI of our Ref. 23 is missing a factor 1/8. The orbital
  paramagnetism in the vicinity of the magic angle is thus about 25 times
  larger than the diamagnetism of the core electrons. But we also note that the
  calculation represents an average value and the details of the band structure
  of the lowest band could still lead to a substantially larger paramagnetic
  response.}\BibitemShut {Stop}%
\bibitem [{Not()}]{Notation}%
  \BibitemOpen
  \href@noop {} {}\bibinfo {note} {The notational swicth is intentional:
  $\ket{n,\bm k} $ is the Bloch eigenstate for {\em band} index $n$, not to be
  confused with the Bloch state for {\em orbital} index $n$, $\ket{\bm k,n}
  $.}\BibitemShut {Stop}%
\end{thebibliography}
\end{document}